\documentclass[a4paper,fleqn,usenatbib]{mnras}

\usepackage{newtxtext,newtxmath}

\usepackage[T1]{fontenc}
\usepackage{ae,aecompl}
\usepackage{multirow}
\usepackage{flushend}

\usepackage{graphicx}	
\usepackage{amsmath}	
\usepackage{amssymb}	
\graphicspath{{./figs/}} 

  \title[Phase and amplitude asymmetry of solar quasi-biennial oscillation]{Phase and amplitude asymmetry in the quasi-biennial oscillation of solar H$\alpha$ flare activity}

\author[L. H. Deng et al.]{
L. H. Deng,$^{1,2,4,5}$
X. J. Zhang,$^{4,6}$
G. Y. Li,$^{1}$
H. Deng$^{3,5}$
and F. Wang$^{2,3,4,5}$\thanks{E-mail: fengwang@gzhu.edu.cn (FW)}
\\
$^{1}$Chongqing University of Arts and Sciences, Chongqing~402160, P.R.~China\\
$^{2}$Center For Astrophysics, Guangzhou University, Guangzhou~510006, P.R.~China\\
$^{3}$Yunnan Observatories, Chinese Academy of Sciences, Kunming~650216, P.R.~China\\
$^{4}$College of Science, China Three Gorges University, Yichang~443002, P.R.~China\\
$^{5}$Key Laboratory of Geospace Environment, Chinese Academy of Sciences, University of Science \& Technology of China, Hefei~230026, P.R.~China\\
$^{6}$CAS Key Laboratory of Solar Activity, National Astronomical Observatories, Beijing~100012, P.R.~China\\
$^{7}$University of Chinese Academy of Sciences, Beijing~100049, P.R.~China}
\date{Accepted 2019 June 11. Received 2019 May 10; in original form 2018 April 26}

\pubyear{2019}

\begin{document}
\label{firstpage}
\pagerange{\pageref{firstpage}--\pageref{lastpage}}
\maketitle

\begin{abstract}
Quasi-biennial oscillation (QBO) of solar magnetic activities is intrinsic to dynamo mechanism, but still far from fully understood. In this work, the phase and amplitude asymmetry of solar QBO of H$\alpha$ flare activity in the northern and southern hemispheres is studied by the ensemble empirical mode decomposition, the cross-correlation analysis, and the wavelet transform technique. The following results are found: (1) solar QBO of H$\alpha$ flare index in the two hemispheres has a complicated phase relationship, but does not show any systematic regularity; (2) the solar cycle mode of solar H$\alpha$ flare index in the northern hemisphere generally leads that in the southern one by 9 months for the time interval from 1966 to 2014. The possible origin of these results is discussed.
\end{abstract}





\section{Introduction}

%
%

The Sun exhibits a quasi-periodic behavior in its magnetic activity, which contains formation of sunspots, faculae, flares, prominences, and other eruptive phenomena. Long-term variation studies of solar magnetic activities revealed that they have a complex dependence on timescales ranging from several seconds to thousands of years and potentially up to millions of years \citep{2004SoPh..221..337M,2017LRSP...14....3U}. In the past few decades, several quasi-periodicities that are shorter and longer than eleven years have been recognized by using a variety of solar activity indicators \citep{2012JASTP..89...48S,2013AstL...39..729P,2014SSRv..186..359B}. The most prominently recognized periodicities shorter than the Schwabe cycle are the quasi-biennial oscillation (QBO) concentrated around 2 years (between 1 and 3 years), whose physical origin may be related to the dynamic process in the solar tachocline \citep{2002A&A...394..701K,2007AdSpR..40.1006O,2007MNRAS.374..282F,2011ApJ...731...30K,2018ApJ...857..113K}. It should additionally be emphasized that periodicities in the range considered here have also been referred to as the mid-term quasi-periodicities (MTQPs, roughly between 1 and 2 years), which were defined by \cite{2000AdSpR..25.1939M}, \cite{2003SoPh..212..201M}, and \cite{2004SoPh..221..337M}.

Numerous studies uncovered that solar QBO could be visible in various manifestations of solar magnetism \citep{1979Natur.278..146S,2009A&A...502..981V} as well as in the interplanetary parameters \citep{2005JGRA..110.1108K,2012ApJ...749..167L,2012SoPh..280..623K}. The QBO was first found in geomagnetic activity index \citep{1972JGR....77.4209F,1975JGR....80.4681D} and auroral activity \citep{1983JGR....88.6310S} at varying levels of significance at different levels. They were also found in the solar wind speed \citep{1994GeoRL..21.1559R,1995GeoRL..22.1845S,1995GeoRL..22.3001P}, cosmic rays \citep{1996SoPh..167..409V}, coronal holes \citep{1992Natur.360..322M}, and solar rotation speed around the tachocline \citep{2000Sci...287.2456H}. The QBO appears to be ubiquitous, but has stochastic characteristics, such as temporal intermittency and variable timescales \citep{2013ApJ...768..188C,2016ApJ...818..127G,2017JGRA..122.5043O}.  Their amplitude are modulated by the 11-year solar cycle, being particularly strong around the maximum phase of solar cycle \citep{2012ApJ...749...27V}. These studies have showed that the QBO is related to the dynamic variation of the solar dynamo process and the emergence of magnetic flux.

Presently, the physical nature of solar QBO is not yet fully understood, but it is believed to be intrinsic to the dynamo mechanism \citep{2003SoPh..212..201M,2005A&A...438..349K,2014SoPh..289..707C}.  Usually, solar QBO is associated with the following probable mechanisms: (i) the quasi-two-year impulse of shear waves could bring out the quasi-two-year periodicity of the photospheric differential rotation \citep{1995SoPh..157...31P}, but this mechanism can work only around the solar minima; (ii) the presence of two types of dynamo actions operating at different depths, one near the top of the layer extending from the surface down to 5\% below it and the other one seated at the base of the convection zone \citep{1998ApJ...509L..49B}; (iii) the spatio-temporal fragmentation of the magnetic Reynolds numbers occurs at the base of the convection zone \citep{2000A&A...363L..13C}; (iv) the beating between a dipole and quadrupole magnetic configuration of the dynamo \citep{2013ApJ...765..100S}; and (v) the instability of magnetic Rossby waves in the solar tachocline \citep{2003MNRAS.345..809L,2010ApJ...724L..95Z,2015NatCo...6E6491M}.

The phase and amplitude asymmetry of solar magnetic activities are an important topic of cyclic behavior that could inform modelers about the relative importance of possible mechanisms that participate in hemispheric coupling \citep{2014SSRv..186..251N,2017ApJ...835...84S}. Several studies examined the north-south asymmetry of different solar activity indicators \citep{2015A&A...582A...4J,2017A&A...603A.109B,2017RAA....17...40L,2018ApJ...855...84X}, and found that solar magnetic fields may be generated independently in the northern and southern hemispheres. \cite{2004ARep...48..678B,2011NewA...16..357B} and \cite{2008SoPh..247..379B} found that solar QBO in the asymmetric time series is even more pronounced and better synchronized than QBO in the indices themselves, and there is a negative correlation between the QBO power and the north-south asymmetry. To the best of our knowledge, the phase and amplitude asymmetry of hemispheric QBO and their potential connection to the 11-year solar cycle mode (SCM) are rarely investigated.


With the hope to add more information on the spatio-temporal distribution and the underlying processes of solar QBO in the northern and southern hemispheres, we focus on H$\alpha$ flare activity over solar cycles 20-24, provided by the Kandilli Observatory of Bogazici University, through the ensemble empirical mode decomposition (ensemble EMD), cross-correlation analysis, and wavelet transform techniques. The observational data and the analysis approaches are briefly introduced in the next Section. In Section 3 the statistical analysis results for the phase and amplitude asymmetry of solar QBO are presented. Finally the conclusions and discussions are given in the last Section 4.

%
%
%

\section{Data and Method}

\subsection{Solar Flare Activity}
\label{sec:data} 

Solar flare activity shown by chromospheric H$\alpha$ observations was described with the flare index, whose concept was first proposed by \cite{1952BAICz...3...52K}. The monthly values of H$\alpha$ flare index, which was considered to be basically proportional to the magnetic energy emitted by solar flares \citep{2003SoPh..214..375O}, were available at the website of the Kandilli Observatory of Bogazici University\footnote{http://www.koeri.boun.edu.tr/astronomy}. Monthly values of H$\alpha$ flare index in the northern and southern hemispheres, during the time interval from 1966 January to 2014 December, are displayed in Figure 1.

\begin{figure}
   \includegraphics[width=\columnwidth]{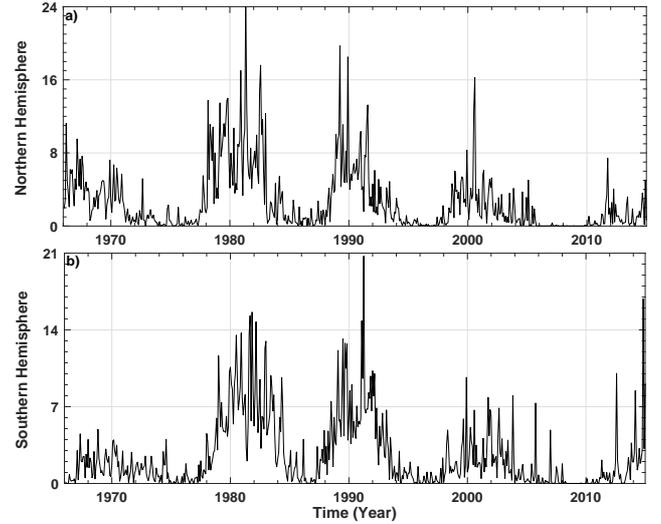}
    \caption{Monthly values of solar H$\alpha$ flare index in the northern (upper panel) and southern (lower panel) hemispheres for the time interval from 1966 January to 2014 December.}
    \label{fig:figure1}
\end{figure}

\subsection{Empirical Mode Decomposition}

Solar QBO could not be analyzed without preparatory filtration, several time-frequency analysis approaches were applied by different authors. Twenty years ago, \cite{1998RSPSA.454..903E} introduced the EMD technique to deal with the nonlinear and non-stationary time series, and this powerful technique has been widely applied in many scientific fields \citep{2004ApJ...614..435T,2010PPCF...52l4009N,2012ApJ...749...27V,2012ApJ...749..167L,2015MNRAS.451.4360K,2016A&A...592A.153K}. 

The key idea of this method is to decompose a complicated data set into a finite and usually small number of intrinsic mode functions (IMFs), representing different oscillations at a local level \citep{2004ApJ...614..435T,2016A&A...589A..56K,2016AJ....151...76X}.  The characteristic timescales of the IMFs extracted by the EMD are based on the local distance between two successive extrema, that is, an IMF represents a hidden oscillation mode, locally defined and thus not-stationary. In a word, the EMD technique is based on the local characteristics of the data set and have a posteriori adaptive basis, it is thus applicable to investigate the periodic variations of the non-linear and non-stationary time series.

A given signal $x(t)$ can be obtained through the reconstruction of its different IMFs with a residual component $r(t)$.

\begin{equation}
x(t)=\sum_{i=1}^{n}IMF_{i}(t)+r(t)
\end{equation}

where $n$ is the total number of IMFs. To perform the decomposition, EMD demands that the signal has at least one maximum and one minimum. Once the maxima and minima are located, the upper and lower envelopes of the time series could be constructed by using cubic spline interpolation. Based on the sifting procedure, the IMFs can be identified if the number of extrema and zero crossings differ at most by one, and at any stage the mean of the envelope defined by local maxima and minima is zero.

For a given time series $x(t)$, the EMD method contains the following steps:

Step(a) Identify total number of extrema in the function.

Step(b) Shifting procedure: (b.1) Using interpolation construct the upper $up_i(t)$ and lower envelope $low_i(t)$ for $i^{th}$ iteration. (b.2) Calculate the envelope mean: $m_i(t)=(up_i(t) + low_i(t))/2$. (b.3) Obtain the signal residue $r_i(t)=x_i(t)-m_i(t)$. (b.4) If $r_i$ satisfies the IMF condition, assign $IMF_j(t)=r_i(t)$ for $j^{th}$ IMF and then update $x(t)$ as $x_{update}(t)=x(t)-IMF_j(t)$ and go to step (a). (b.5) Else go to step (a).

Step(c) Stopping criteria: (c.1) Iterate over residue and check the extrema of $r_i(t)$ after each sifting procedure. (c.2) Stop sifting if extrema is one or $r(t)$ becomes
a constant or a monotonic function.

\begin{figure}
    \includegraphics[width=\columnwidth]{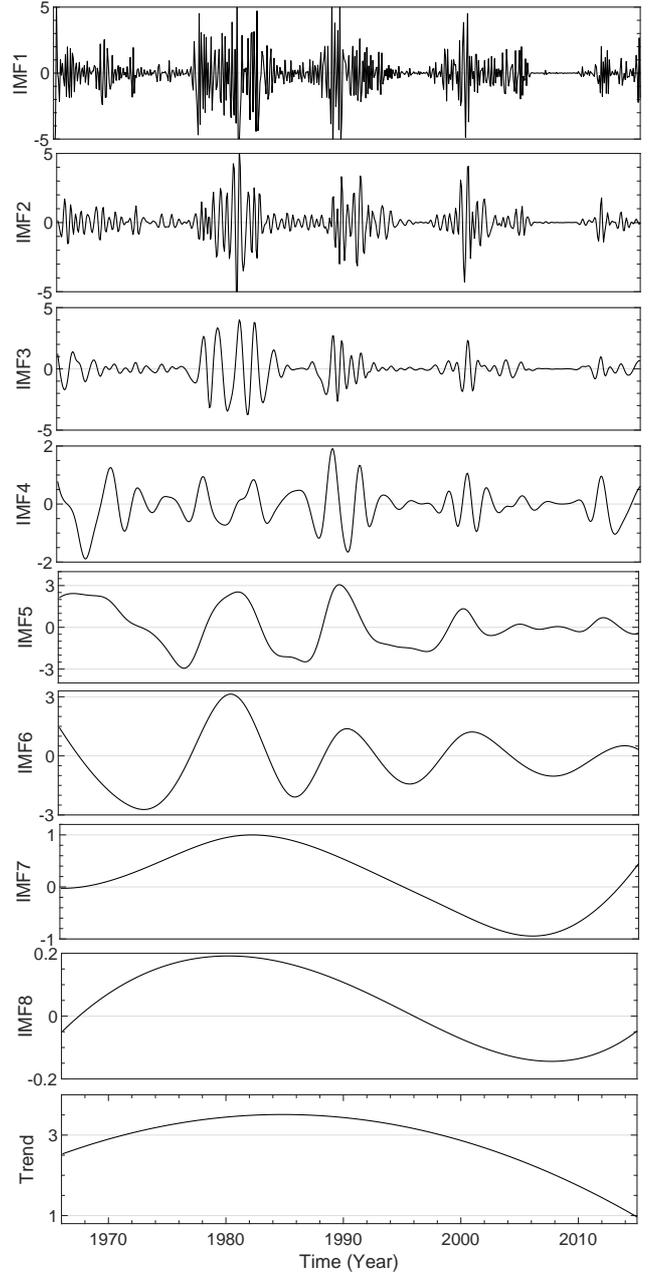}
    \caption{The average IMFs of solar H$\alpha$ flare index in the northern hemispheres. IMFs 1-8 and the trend are shown correspondingly in the panels, ranking from the top to the bottom, respectively.}
    \label{fig:figure2}
\end{figure}

\begin{figure}
    \includegraphics[width=\columnwidth]{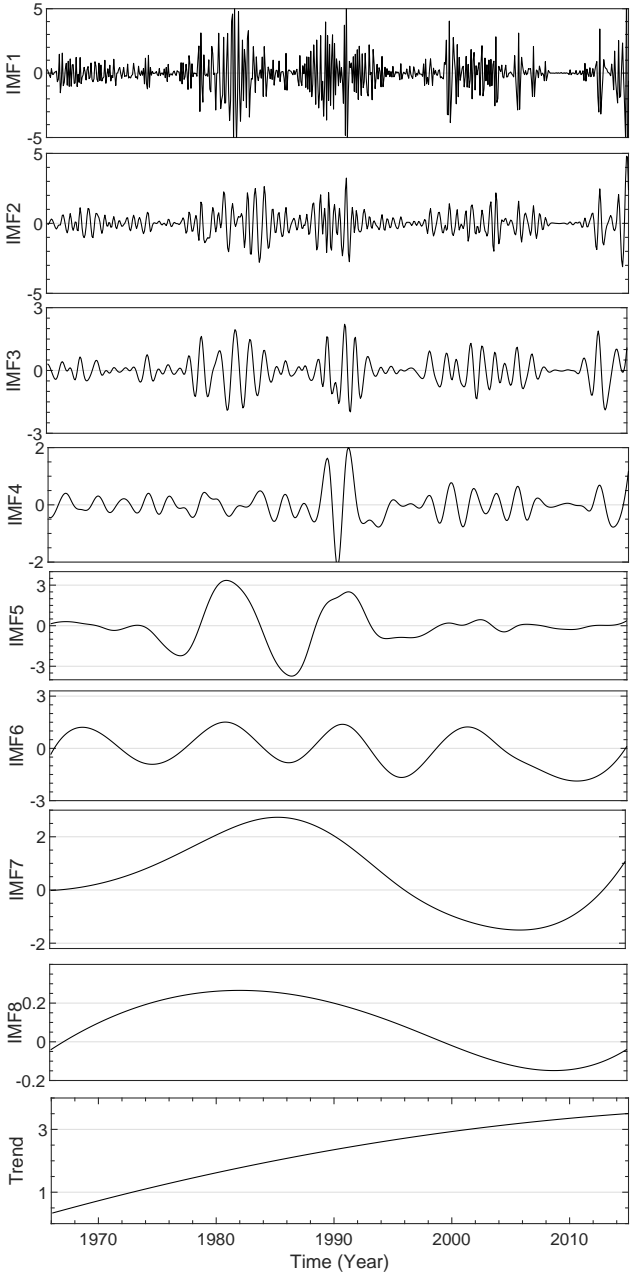}
    \caption{The average IMFs of solar H$\alpha$ flare index in the southern hemispheres. IMFs 1-8 and the trend are shown correspondingly in the panels, ranking from the top to the bottom, respectively.}
    \label{fig:figure2}
\end{figure}

\subsection{Ensemble Empirical Mode Decomposition}

The major shortcoming of the EMD technique is the frequent appearance of mode mixing, which may lead to serious aliasing in the time-frequency distribution, and also make the physical meaning of an IMF unclear \citep{2009AADA...1...1}. The mode mixing phenomenon occurs when the oscillations with disparate time scale are preserved in one IMF, or the oscillations with the same time scale are sifted into different IMFs.

To overcome the mode mixing problem, an improved method was developed, the ensemble EMD (EEMD) which defines the IMF components as the mean of an ensemble, each consisting
of the signal plus a white noise of finite amplitude. Here, the term $mean$ is not the mean of the white noise, but the average of the total $i$th IMF related to $N$ (the ensebmle number) trials. The steps of EEMD algorithm are as follows:

Step(a): Add a white noise series to the given time series.

Step(b): Decompose the data with added white noise into IMFs (EMD previously explained).

Step(c): Repeat step (1) and step (2) a certain number of iterations.

Step(d): Obtain the (ensemble) means of corresponding IMFs of the decompositions as the final result.

White noise would populate the whole time-frequency space uniformly with the constituting components at different scales. Although each individual trial may produce very noisy results, the noise in each trial is canceled out in the ensemble mean of all trials.

The results achieved by the EEMD depend on the choice of the ensemble number ($N$) and the amplitude of added white noise ($A$). Within a certain window of noise amplitude, the sensitivity of the decomposition of data using the EEMD method to the amplitude of noise is often small. In this study, noise with a standard deviation of 0.02, 0.04, ..., and 0.4 (step size is 0.02, total 20 cases) is added, while the ensemble size is set to 100 in each case.

\begin{figure}
    \includegraphics[width=\columnwidth]{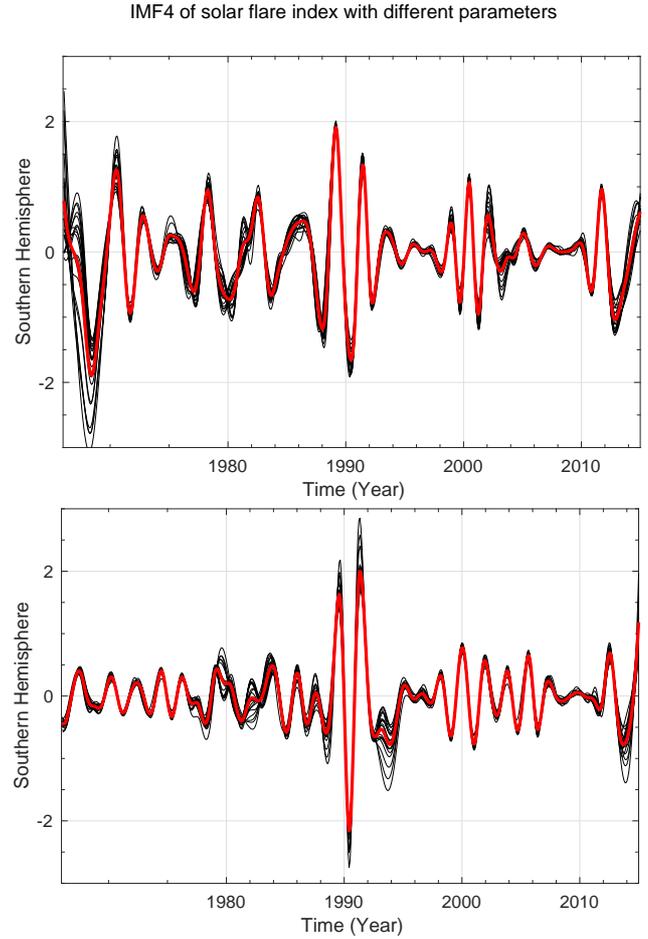}
    \caption{IMF4 of solar flare index in the northern (upper panel) and southern (lower panel) hemispheres, respectively. In each panel, black lines (number: 20) correspond to EEMD algorithm with added noise of standard deviation from 0.02 to 0.40 with a step of 0.02, and the bold red line is the average IMF4 of 20 time series. The ensemble number for each case is 100.}
    \label{fig:figure2}
\end{figure}

\begin{figure}
    \includegraphics[width=\columnwidth]{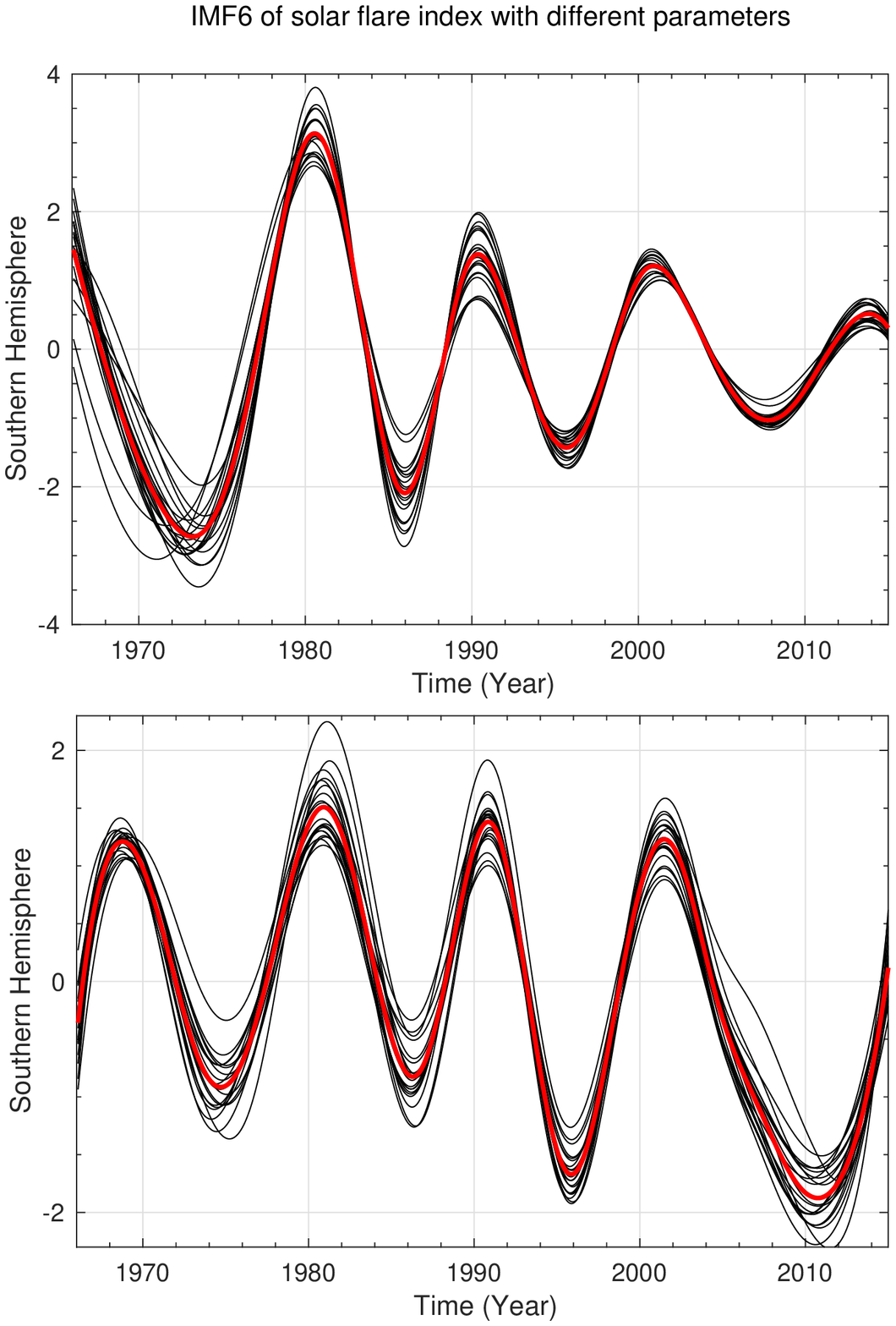}
    \caption{IMF6 of solar flare index in the northern (upper panel) and southern (lower panel) hemispheres, respectively. In each panel, black lines (number: 20) correspond to EEMD algorithm with added noise of standard deviation from 0.02 to 0.40 with a step of 0.02, and the bold red line is the average IMF6 of 20 time series. The ensemble number for each case is 100.}
    \label{fig:figure2}
\end{figure}

\section{Analysis Results}

By applying the EEMD technique, eight IMFs and a trend are obtained for each considered data set. As the amplitude of added noise is changed 20 times, so we have 20 time series for each IMF. At a certain timescale, the final IMF is calculated as the average of 20 time series. Figure 2 and Figure 3 display the average IMFs and the trend of solar H$\alpha$ flare index in the northern and southern hemispheres, respectively.

To show the decomposition results between cases of different levels of added noise, the IMF4 and IMF6 of solar H$\alpha$ flare index in the two hemispheres are taken as an example. Figure 4 displays the temporal variations of IMF4 in the northern and southern hemispheres, respectively. In each panel, 20 black lines correspond to EEMD algorithm with added noise of standard deviation from 0.02 to 0.40, and the bold red line is the average IMF4 of 20 time series. Similar to Figure 4, Figure 5 shows the temporal variabilities of IMF6 in the two hemispheres. Clearly, the synchronization between cases of different levels of added noise is remarkably good. Therefore, the EEMD provides a sort of uniqueness and robustness result that the original EMD usually could not, and it increase the confidence of the decomposition.

\begin{figure}
    \includegraphics[width=\columnwidth]{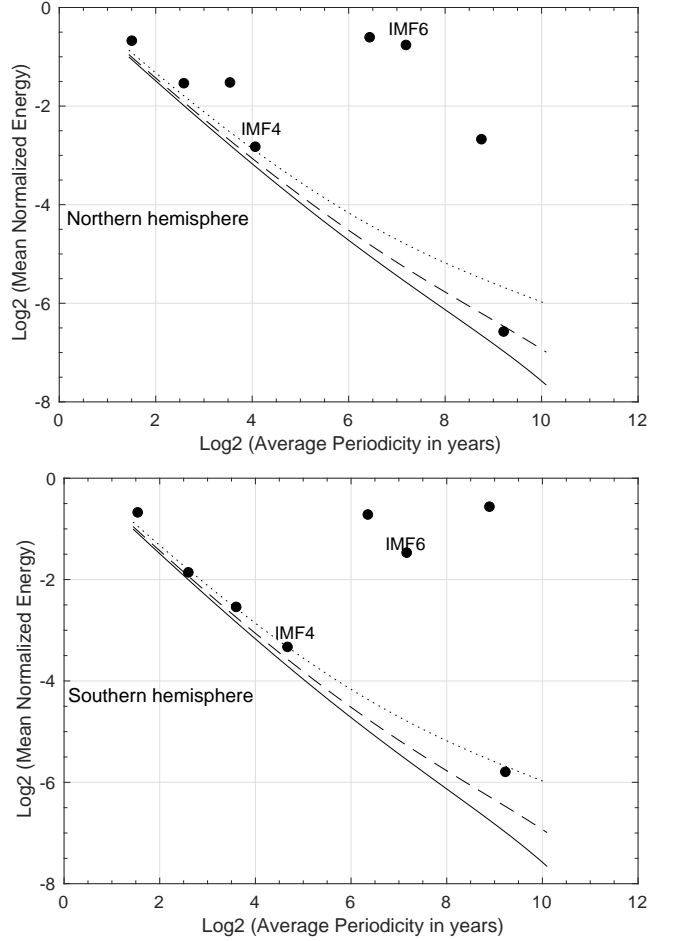}
    \caption{Statistical significance test of the eight IMFs of solar H$\alpha$ flare index in the northern (upper panel) and southern (lower panel) hemispheres, respectively. For each panel, the solid, dashed, and dotted lines represent the spread lines at the 90th, 95th, and 99th percentile, respectively.}
    \label{fig:figure2}
\end{figure}

In our analysis, the statistical significance of information content for the extracted eigenmodes is estimated by applying the test proposed by \cite{2004RSPSA.460.1597W}. This test is based on the comparison between the IMFs obtained from the signal with the corresponding ones derived from a white noise process. For example, if the energy of the first IMF is $E_1$, then we can obtain the relative energy of the $i$th IMF ($E_i$) using the energy level of IMF1. 

The energy density of the $i$th IMF is defined as:

\begin{equation}
E_{i}=\sum_{j=1}^{N}|\operatorname{IMF}_i(j)|^{2}
\end{equation}

 where $N$ is the number of data points. Based on this equation, we can obtain the relative energy of all IMFs using the energy level of IMF1. Also, the spread function of the different confidence levels of white noise can be calculated. If the energy level of any IMF lies above the spread line, the IMF is statistically significantly at this confidence level. In this work, we select three confidence-limit level: 90th, 95th, and 99th percentile. 

The analysis results from the statistical test performed on the solar H$\alpha$ flare index in the two hemispheres are shown in the Figure 6, where the solid, dashed, and dotted lines represent the confidence levels at the 90th, 95th, and 99th percentile, respectively. As shown in this figure, most of the IMFs are above the 95th percentile spread line and could be considered statistically significant. The eight modes along with their typical periodicities and the period ranges of solar H$\alpha$ flare index in the two hemispheres are collected in Table 1. The inherent problem of the EEMD method is the fact that the periodicities of IMFs are not stationary, so the period range of each IMF is calculated. Here, the period range of each IMF is calculated as the standard deviation of the corresponding periodicities of the 20 time series (for example, as shown in Figures 4 and 5).

\begin{table}
	\centering  
	\caption{The average periodicities and the period ranges (in years) of eight IMFs which are extracted from the solar flare index in the two hemispheres.}.
	\begin{tabular}{ccccc}
		\hline\hline
		          &  Northern Hemisphere   & Southern Hemisphere\\
		\hline
		IMF1	  & 0.2390$\pm$0.0025			          & 0.2450$\pm$0.0080	                      \\
		IMF2  & 0.5052$\pm$0.0071			          & 0.5104$\pm$0.0178			     \\
		IMF3	  & 0.9800$\pm$0.0242	                           & 1.0208$\pm$0.0142	             \\
		IMF4	  & 1.4099$\pm$0.2280	                           & 2.1471$\pm$0.0582		     \\
		IMF5	  & 7.2932$\pm$0.8631			          & 6.8799$\pm$1.5934                      \\
		IMF6	  & 12.291$\pm$0.1425		                   &12.076$\pm$0.1418		     \\
		IMF7  & 36.350$\pm$7.1036	                           &40.074$\pm$	5.1437	             \\
		IMF8	  & 50.025$\pm$1.8436	                           &50.475$\pm$	1.9453		    \\
		\hline
	\end{tabular}
	\label{T_ccx_subregion}
\end{table}

\subsection{Reconstruction of Solar QBO and SCM}

From Figure 6 and Table 1 one can easily see that the IMF6 for each time series represents the $\sim$11-year periodicity, which can be defined as the SCM for further analysis. The average periodicities of the IMF6 are calculated to be $12.291\pm0.1425$ years and $12.076\pm0.1418$ years for solar H$\alpha$ flare index in the northern and southern hemispheres, respectively.

The average periodicity of IMF4 for each data set is associated with the typical timescales between 1 and 3 years, one is $1.4099\pm0.2280$ years, the other is $2.1471\pm0.0582$ years, so the IMF4 of H$\alpha$ flare activity in the two hemispheres could be considered as the solar QBO.

Here, it should be pointed out that why IMF3 is not taken as a part of solar QBO. The typical periodicity of IMF3 is $0.9800\pm0.0242$ years in the northern hemisphere and $1.0208\pm0.0142$ years in the southern hemisphere. On the one hand, a large fraction of IMF3 is outside the QBO definition. On the other hand, this IMF is most likely related to the annual-variation signal. The one-year periodicity has been found in many solar activity indicators, but its origin is still doubtful. It is difficult to rule out the possibility that this periodicity is not due to the Earth's orbital revolution \citep{2009SoPh..257...61J}. Up to now, there has been no quantitative analysis about the effect of Earth's helio-latitude on the measurement of the Sun, and the physical origin of the one-year periodicity is an open issue.

\subsection{Phase Relationship of Hemispheric QBO}

\begin{figure}
    \includegraphics[width=\columnwidth]{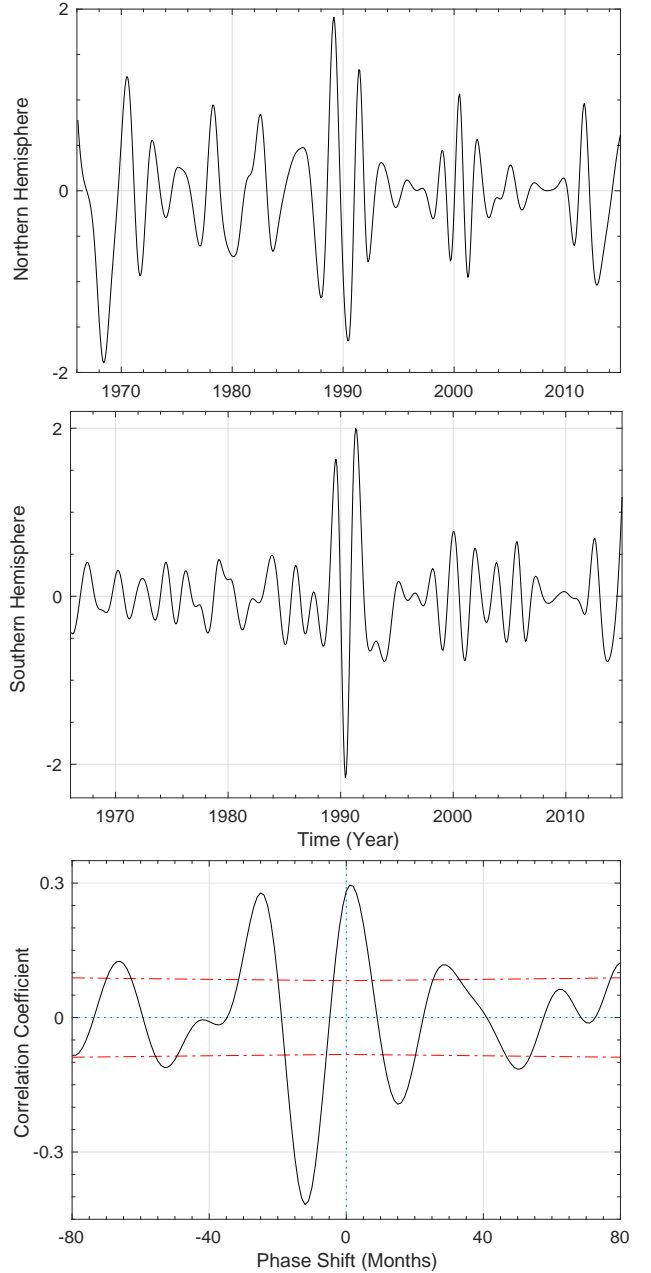}
    \caption{Top panel:  the QBO (IMF4) of solar flare activity in the northern hemisphere. Middle panel: similar to the top panel, but for the southern hemisphere. Bottom panel: the cross-correlation analysis results of the hemispheric QBO, and the red dash-dotted lines are the 95\% confidence levels.}
    \label{fig:figure6}
\end{figure}

There are many statistical analysis techniques to investigate the phase relationship of solar time series. Generally speaking, the series of auto-correlation coefficients represents the correlation between the successive data of a single time series. The cross-correlation analysis could be applied to determine the degree of fit between two time series. To investigate the phase relationship of hemispheric QBO of solar flare activity in the two hemispheres, the cross-correlation analysis method is chosen in this work.

The top and middle panels of Figure 7 display the temporal variabilities of solar QBO of H$\alpha$ flare activity in the northern and southern hemisphere, respectively. From this figure one can easily see that the temporal variabilities of solar QBO in the two hemispheres behave differently during the considered time interval, suggesting that solar QBO should be asynchronous in the northern and southern hemispheres.

The bottom panel of Figure 7 shows the results of the cross-correlation analysis of the hemispheric QBO of solar H$\alpha$ flare activity with the phase lags between -80 and 80 months. The abscissa indicates the phase shift of the northern hemisphere with respect to the southern one along the calendar-time axis, with positive values representing forward shifts (i.e., the northern hemisphere begins earlier in time). The red dash-dotted lines are the 95\% confidence levels, which are calculated by the standard MATLAB code for cross-correlation analysis. Here, the  relative phase shifts are only chosen from -80 to 80 months, so most of the local peaks are above the 95\% confidence levels.

When there is no phase shift between the two, the correlation coefficient is 0.28, indicating that solar QBO in the two hemispheres are positive correlation. When the relative phase shifts between the two are -53, -12, 15, and 50 months, the values of the correlation coefficient reach local minima of -0.11, -0.42, -0.19, and -0.11, respectively. The average interval between each two neighboring local minima is 34$\pm$7 months (2.83$\pm$0.58 years). When the phase shifts are -66, -25, 1, and 29 months, the values of the correlation coefficient peak at local maxima of 0.13, 0.28, 0.30, and 0.12, respectively. The average interval between each two neighboring local maxima is 32$\pm$8 months (2.67$\pm$0.21 years). Therefore, solar QBO of H$\alpha$ flare activity in the two hemispheres have a complicated phase relationship, but does not show any systematic regularity.

\subsection{Phase Relationship of Hemispheric SCM}

\begin{figure}
    \includegraphics[width=\columnwidth]{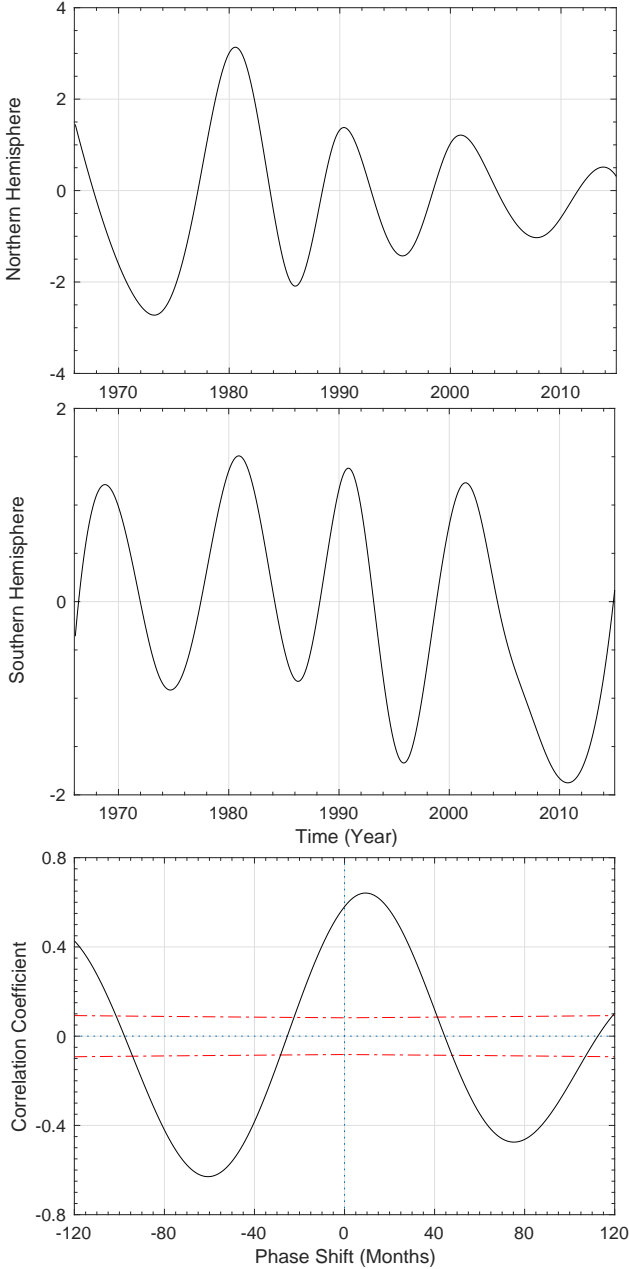}
    \caption{Top panel:  the SCM (IMF6) of solar flare activity in the northern hemisphere. Middle panel: similar to the top panel, but for the southern hemisphere. Bottom panel: the cross-correlation analysis results of the hemispheric SCM, and the red dash-dotted lines are the 95\% confidence levels.}
    \label{fig:example_figure}
\end{figure}

The top and middle panels of Figure 8 display the temporal variabilities of the SCM of solar H$\alpha$ flare activity in the northern and southern hemisphere, respectively. From this figure one can easily see that the temporal variabilities of the SCM in the two hemispheres behave differently during the considered time interval, implying a slight decoupling between the two hemispheres. 

The bottom panel of Figure 8 shows the results of the cross-correlation analysis of the hemispheric SCM with the phase lags between -120 and 120 months. Their phase difference is found to be  9 months where the correlation coefficient has a largest value (0.64), which is above the 95\% confidence levels shown by the red dash-dotted lines. It is worthy of note that the phase difference obtained by us is slightly larger than the results given by previous authors, such as 5-7 months by studying the sunspot areas \citep{2015NewA...39...55R} and flare index \citep{2017JSWSC...7A..34D}. 

For all we know, the above-mentioned studies only smoothed the solar time series with 13 months, but did not consider the different periodic components  that are responsible for the hemispheric phase difference. For example, \cite{2010MNRAS.401..342L} found that the high-frequency components of hemispheric flare activity exhibit an asynchronous behavior with strong phase mixing, but the low-frequency components, corresponding to the timescales around the Schwabe cycle (8-12 years), display strong synchronous behavior with coherent phase angles. Therefore, the temporal variability of the SCM in the northern hemisphere begins 9 months earlier than that in the southern one during the time interval from 1966 to 2014.

\subsection{Amplitude Asymmetry of Hemispheric QBOs}

\begin{figure}
    \includegraphics[width=\columnwidth]{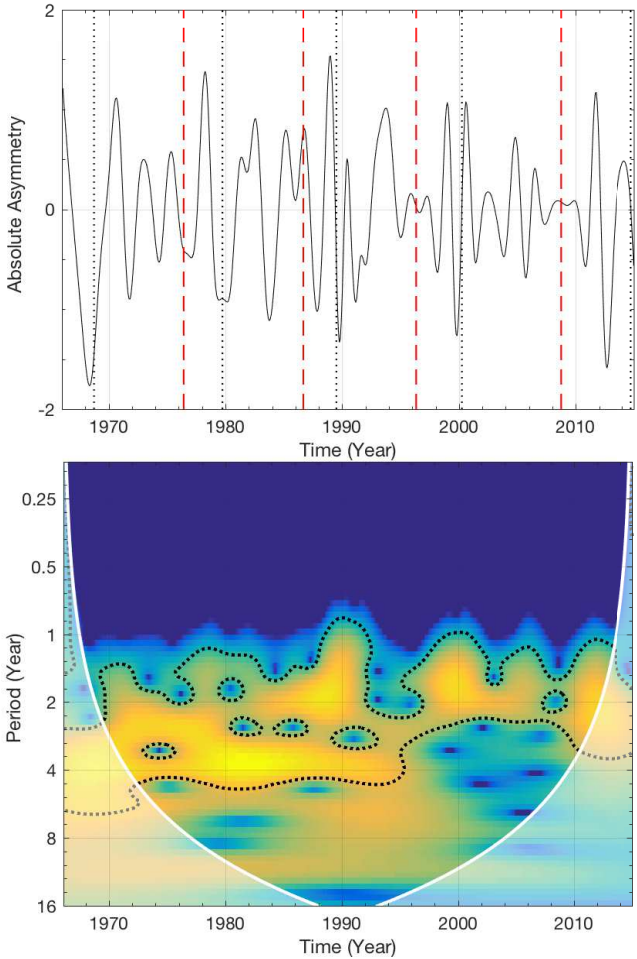}
    \caption{Upper panel: the time series of $A_{ns}$ for the hemispheric QBO of solar H$\alpha$ flare activity. Lower panel: wavelet power spectra of the $A_{ns}$ time series for the hemispheric QBO of solar H$\alpha$ flare activity. The black dotted contours outline the regions where the confidence level is above 95\%, and the white solid line represents the cone of influence (COI), in which the wavelet transform suffers from the edges effects. }
    \label{fig:example_figure}
\end{figure}

Since 1955, the north-south asymmetry (i.e., the hemispheric asymmetry) of solar magnetic activities has been widely studied by introducing the normalized asymmetry index $N_{ns}$ \citep{1955MNRAS.115..398N} and the absolute asymmetry index $A_{ns}$ \citep{2005A&A...431L...5B}. Here, the time series of $N_{ns}$ is defined as the $(Q_n-Q_s)/(Q_n+Q_s)$, where $Q_n$ and $Q_s$ stand for the values of the considered solar activity index corresponding to the northern and southern hemispheres, respectively. The time series of $A_{ns}$, $Q_n-Q_s$, is calculated as the absolute difference between the amplitude indices $Q_n$ and $Q_s$.

 \cite{2005A&A...431L...5B} pointed out that the statistical analysis results obtained by the application of the power spectrum analysis to the time series of $N_{ns}$ are misleading. In this work, the definition of the time series of $A_{ns}$ is thus applied to investigate the periodic variation of hemispheric QBO of solar H$\alpha$ flare activity. The time series of $A_{ns}$ for hemispheric QBO is shown in  the upper panel of Figure 9. From this figure one can see that the hemispheric QBO of solar H$\alpha$ flare activity exhibits a strong asymmetry around the times of solar maximum (indicated by red dashed lines). 
 
 To perform the periodic behavior of the hemispheric asymmetry of solar QBO of H$\alpha$ flare activity, the wavelet transform software proposed by \cite{1998BAMS...79...61T} is chose to analyze the localized variations in the time-frequency space, and the analysis results are shown in the lower panel of Figure 9. In our analysis, we choose the Morlet function as the mother function, and the red-noise significance test is performed. 
 
 For the $A_{ns}$ time series of the hemispheric QBO, the periodic scales ranging from 1 to 3 years are most dominant during the solar maxima of cycles 21-24, especially in cycles 21 and 22, but they are absent around the times of cycle minimum. Our results are agreement with and further enhance the previous results given by other authors. For instance, \cite{2004ARep...48..678B,2011NewA...16..357B} and \cite{2008SoPh..247..379B} found that solar QBO in the asymmetric time series are even more pronounced and better synchronized than QBO in the indices themselves, and there is a negative correlation between the QBO power and the north-south asymmetry.

\section{Conclusions and Discussions}

Using the solar H$\alpha$ flare index in the northern and southern hemispheres from 1966 January to 2014 December, we presented the statistical analyses of the phase and amplitude asymmetry of solar oscillation modes at different timescales. Firstly, the EEMD technique was applied to extract the IMFs of solar H$\alpha$ flare activity and to reveal the periodicities. Then, the phase asynchronism of the hemispheric QBO and SCM were separately studied by the cross-correlation analysis method. And finally, the wavelet transform analysis was applied to study the periodic variation of the north-south asymmetry of the hemispheric QBO. 

By using the EEMD method, the time series of solar H$\alpha$ flare index in the two hemispheres are decomposed into eight IMFs and a trend. The IMF6 for each data set represents the $\sim$11-year periodicity, which could be defined as the SCM. The average periodicity of IMF4 for each data set is associated with the typical timescales of 1-3 years, so the IMF4 in each hemisphere is considered as the solar QBO.

Based on the cross-correlation analysis, solar QBO in the two hemispheres has a complicated phase relationship, but does not show any systematic regularity. However, the SCM in the northern hemisphere begin 9 months earlier than that in the southern one during the considered time interval. That is, the phase relationship of the hemispheric QBO differs from that of the hemispheric SCM. The former one is very complex, while the latter one is very simple. They are not correlated with each other, so they may have either different physical origins or the similar process with different parameters. A possible reason is that the magnetic field strength and the differential rotational parameters in the two hemispheres of the Sun vary in time depending on the amplitude and phase in a certain cycle, which brings out different periodicities and growth rates of solar magnetic fields in the northern and southern hemispheres.

Previous studies have often revealed the existence of the phase and amplitude asymmetry of solar activity indicators in the two hemispheres, and there are several possibly theoretical explanations based on the solar dynamo theories. For example, \cite{2009RAA.....9..115G} demonstrated that the randomness in the dynamo processes can lead to the amplitude of the poloidal fields stronger in one hemisphere than the other one. The asymmetric polar-field reversals should be related to the hemispheric asynchronism of solar activities, which has been found by \cite{2013ApJ...763...23S}. \cite{2015ApJ...806..169B} found that the effect of greater inflows into the active region in one hemisphere can make the activity level in that hemisphere larger compare to the other one.\cite{2017ApJ...835...84S} used the mean field theory to study the intercorrelation between the dipole- and quadrupole-type components of solar magnetic fields. They found that two different attractors may exist in the solar activity cycle, and this can be applied to explain the phase asynchronism of solar activities and polar-field reversal. Based on an updated Babcock-Leighton-type dynamo model, \cite{2018A&A...618A..89S} showed that the absolute hemispheric asymmetry of solar magnetic activity could be naturally explained by the superposition of an excited dipolar mode and a linearly damped, but randomly excited quadrupolar mode. We wish that, in the near future, more observational data sets driven numerical simulations required to better understand and reveal the physical process of the phase and amplitude asymmetry of solar QBO.

\section*{Acknowledgements}


Solar H$\alpha$ flare index used in this work were calculated by T. Atac and A. Ozguc from Bogazici University Kandilli Observatory, Istanbul, Turkey. This work is supported by the Joint Research Fund in Astronomy (Nos. U1631129, U1831204) under cooperative agreement between the National Natural Science Foundation of China (NSFC) and Chinese Academy of Sciences (CAS), the National Natural Science Foundation of China (No. 11873089), the Youth Innovation Promotion Association CAS, the CAS ``Light of West China'' Program, the Yunnan Key Research and Development Program (2018IA054), the open research program of CAS Key Laboratory of Solar Activity (Nos. KLSA201807), the Key Laboratory of Geospace Environment, CAS, University of Science \& Technology of China, and the major scientific research project of Guangdong regular institutions of higher learning (2017KZDXM062). Last, but not least, the authors wish to express their gratitude to the anonymous referee for constructive suggestions on our manuscript that greatly improved the content and presentation of this work.




\bibliographystyle{mnras}
\bibliography{eemd_sfi} 

\bsp	
\label{lastpage}
\end{document}